\documentclass[conference,a4paper]{APSIPA2021}
\usepackage{amsmath}
\usepackage{graphicx}
\usepackage{multirow}
\usepackage{threeparttable}
\usepackage{color}

\usepackage{amssymb}
\usepackage{subcaption}
\usepackage{cite}
\usepackage{url}

\usepackage{geometry}
\usepackage{fancyhdr}

\fancypagestyle{firststyle}{
  \fancyhf{}
  \fancyhead[C]{2024 Asia Pacific Signal and Information Processing Association Annual Summit and Conference (APSIPA ASC)}
}

\begin{document}

\title{On the Security of Bitstream-level JPEG Encryption with Restart Markers}

\author{
\authorblockN{
Mare Hirose\authorrefmark{1},  Shoko Imaizumi\authorrefmark{1} and Hitoshi Kiya\authorrefmark{2}
}

\authorblockA{
\authorrefmark{1}
Chiba University, Chiba, Japan \\
E-mail: marehirose@chiba-u.jp, imaizumi@chiba-u.jp}

\authorblockA{
\authorrefmark{2}
Tokyo Metropolitan University, Tokyo, Japan\\
E-mail: kiya@tmu.ac.jp}
}

\maketitle
\thispagestyle{firststyle}
\pagestyle{empty}

\begin{abstract}
This paper aims to evaluate \textcolor{black}{the security} of a bitstream-level JPEG encryption method \textcolor{black}{using} restart (RST) markers, \textcolor{black}{where} encrypted image \textcolor{black}{can keep} the JPEG file format with the same file size as \textcolor{black}{non-encrypted image}. Data encrypted \textcolor{black}{using this} method can be decoded without \textcolor{black}{altering} header information by \textcolor{black}{employing} a standard JPEG decoder. \textcolor{black}{Moreover}, the use of RST markers \textcolor{black}{enables the definition of} extended blocks \textcolor{black}{divided} by the markers, so spatially partial encryption and block-permutation-based encryption can be carried out. However, the security of the method was evaluated only \textcolor{black}{with respect to} the key space analysis for brute-force attacks and other limited attacks.
Accordingly, in this paper, we evaluated the security of the method \textcolor{black}{with respect to} robustness against ciphertext-only attacks including state-of-the-art attacks. In experiments, the method is compared with conventional encryption methods, and it is confirmed to be robust against ciphertext-only attacks if parameters used for image encryption are carefully chosen.
\end{abstract}

\section{Introduction}
The use of JPEG images has \textcolor{black}{grown} due to the \textcolor{black}{expansion} of social networking services. These \textcolor{black}{platforms generally restrict the use of various file formats}. In many cases, images have \textcolor{black}{to adhere to} the JPEG standard. \textcolor{black}{Furthermore}, these \textcolor{black}{platforms} are not reliable \textcolor{black}{owing to} incidents\textcolor{black}{, including} information leakage. Accordingly, many encryption methods that \textcolor{black}{keep} the JPEG format have been proposed \textcolor{black}{to date}. Recently, a bitstream-level encryption method for JPEG compression was proposed, and it was \textcolor{black}{confirmed} to outperform conventional encryption methods in some important respects \cite{enc_rst}. However, it has not been verified enough yet \textcolor{black}{with respect to} the robustness of the encryption against various attacks. Accordingly, we aim to evaluate the security of the method against ciphertext-only attacks (COAs) in this paper.

Bitstream-level encryption is one of the types of encryption combined with image compression. Compared with other types, this type of encryption has some advantages. For example, it \textcolor{black}{permits} us not \textcolor{black}{just} to \textcolor{black}{keep} the JPEG file format but also to use standard JPEG encoders. In addition, some methods of this type ensure that the file size remains the same before and after encryption, so encryption can be carried out by hooking images within a transmission channel. One state-of-the-art method \cite{enc_rst} can also generate partially encrypted images, but all other bitstream-level methods cannot.

In \cite{enc_rst}, the restart (RST) marker, \textcolor{black}{placed} at fixed intervals between minimum coded units (MCUs), is used for encryption. This marker contributes to the mixing of encrypted and unencrypted regions. Furthermore, MCUs \textcolor{black}{divided} by RST markers can be defined as extended blocks, and the visibility of the encrypted image can be reduced by permuting the position of these blocks. This block permutation is also expected to enhance security strength.

Accordingly, we focus on the above state-of-the-art method to evaluate the attack resistance of the method against ciphertext-only attacks (COAs) in this paper. Key space analysis, key sensitivity analysis, the non-zero counting attack (NZCA), and histogram analysis are performed, and the relationship between the use of RST markers and the attack resistance is discussed. In experiments, it is demonstrated that the RST marker, used by the method used in image encryption for the first time, can enhance the robustness against \textcolor{black}{a variety of} attacks in general, and the attack resistance of encrypted images depends on the selection of restart interval values.

\section{Related Work}
The encryption method discussed in this paper is a method \textcolor{black}{coupled} with image compression that \textcolor{black}{is able to produce} JPEG files \textcolor{black}{containing obscured} visual information \cite{enc_rst}. \textcolor{black}{Accordingly, conventional} encryption methods \textcolor{black}{coupled} with image compression are \textcolor{black}{reviewed} here. \textcolor{black}{Furthermore}, the \textcolor{black}{structure} of JPEG bitstreams is briefly explained.

\subsection{Combined Use of Encryption and Compression}
\begin{figure}[t]
  \centering
  \begin{minipage}[b]{\linewidth}
    \centering
    \includegraphics[width=\linewidth]{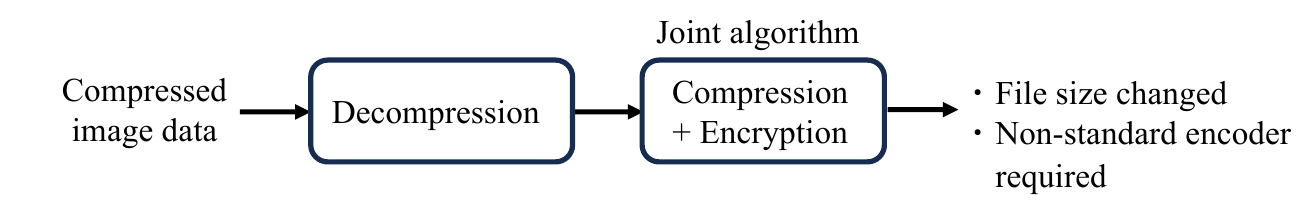}
    \subcaption{Type 3}\label{fig:type3}
  \end{minipage} \\
  \vspace{0.5cm} 
  \begin{minipage}[b]{\linewidth}
    \centering
    \includegraphics[width=\linewidth]{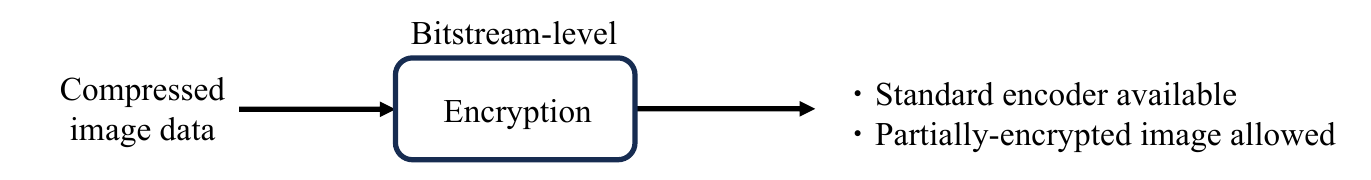}
    \subcaption{Type 4}\label{fig:type4}
  \end{minipage}
  \caption{Types of encryption.}
  \label{fig:encryption}
\end{figure}

Encryption methods \textcolor{black}{coupled} with image compression are \textcolor{black}{categorized} into four types as \textcolor{black}{follows}. Type 1 is compression-then-encryption (CtE) \textcolor{black}{using} standard cryptography, in which images are compressed and then encrypted \textcolor{black}{using} a standard cryptography \textcolor{black}{including} AES (advanced encryption standard). Type 2 is encryption-then-compression (EtC), in which the visual information of images is \textcolor{black}{secured through} a perceptual encryption method, \textcolor{black}{referred to as} compressible encryption, and \textcolor{black}{afterward} the encrypted images \textcolor{black}{undergo compression}\cite{etc1, etc2}. Type 3 combines compression and encryption. For type 3, encryption and compression are \textcolor{black}{processed at the same time} (see Fig. \ref{fig:type3}) \cite{enc1, enc2, enc5, enc7, peiya}. \textcolor{black}{Therefore, when JPEG images are provided before encryption}, the images \textcolor{black}{need to} be decompressed \textcolor{black}{prior to performing} encryption. \textcolor{black}{Furthermore,} standard encoders cannot be used in this framework. Type 4 is CtE \textcolor{black}{using} a bitstream-level encryption method (see Fig. \ref{fig:type4}), in which images \textcolor{black}{undergo compression} and then \textcolor{black}{encryption using} a bitstream-level encryption method \cite{enc3, enc4, enc6, pre, enc_rst}. Type 4 \textcolor{black}{permits} us not just to \textcolor{black}{keep} the JPEG format but also to use
standard JPEG decoders. 

We focus on a state-of-the-art method \cite{enc_rst}, which is a type 4 method for JPEG images. The method enables us to \textcolor{black}{produce} encrypted images with \textcolor{black}{different} levels of visibility, \textcolor{black}{such as} partially encrypted images.

\subsection{JPEG Bitstream and Marker Codes}
\begin{figure}[t]
\begin{center}
\includegraphics[width=80mm]{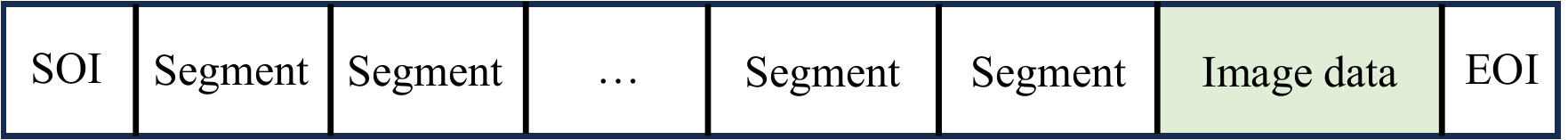}
\end{center}
\caption{\textcolor{black}{Structure} of JPEG bitstream.}
\label{fig:structure}
\vspace*{-3pt}
\end{figure}

Here, we describe the \textcolor{black}{structure} of the JPEG bitstream. Fig. \ref{fig:structure} \textcolor{black}{illustrates} an example of the \textcolor{black}{structure} of JPEG bitstreams \cite{iso}. Codes called SOI and EOI, which respectively signify the beginning and end of a bitstream, are allocated. These are two-byte codes called marker codes. Segments \textcolor{black}{include} information necessary for decoding, such as Huffman tables and quantization tables. Marker codes are also allocated at the beginning of each segment. The first byte of the marker code is fixed at FF$_{(16)}$, and the second byte indicates the type of each marker. The second byte of the marker code distinguishes each segment. Note that FF00$_{(16)}$ is not defined as a marker code.

\begin{figure}[t]
\begin{center}
\includegraphics[width=80mm]{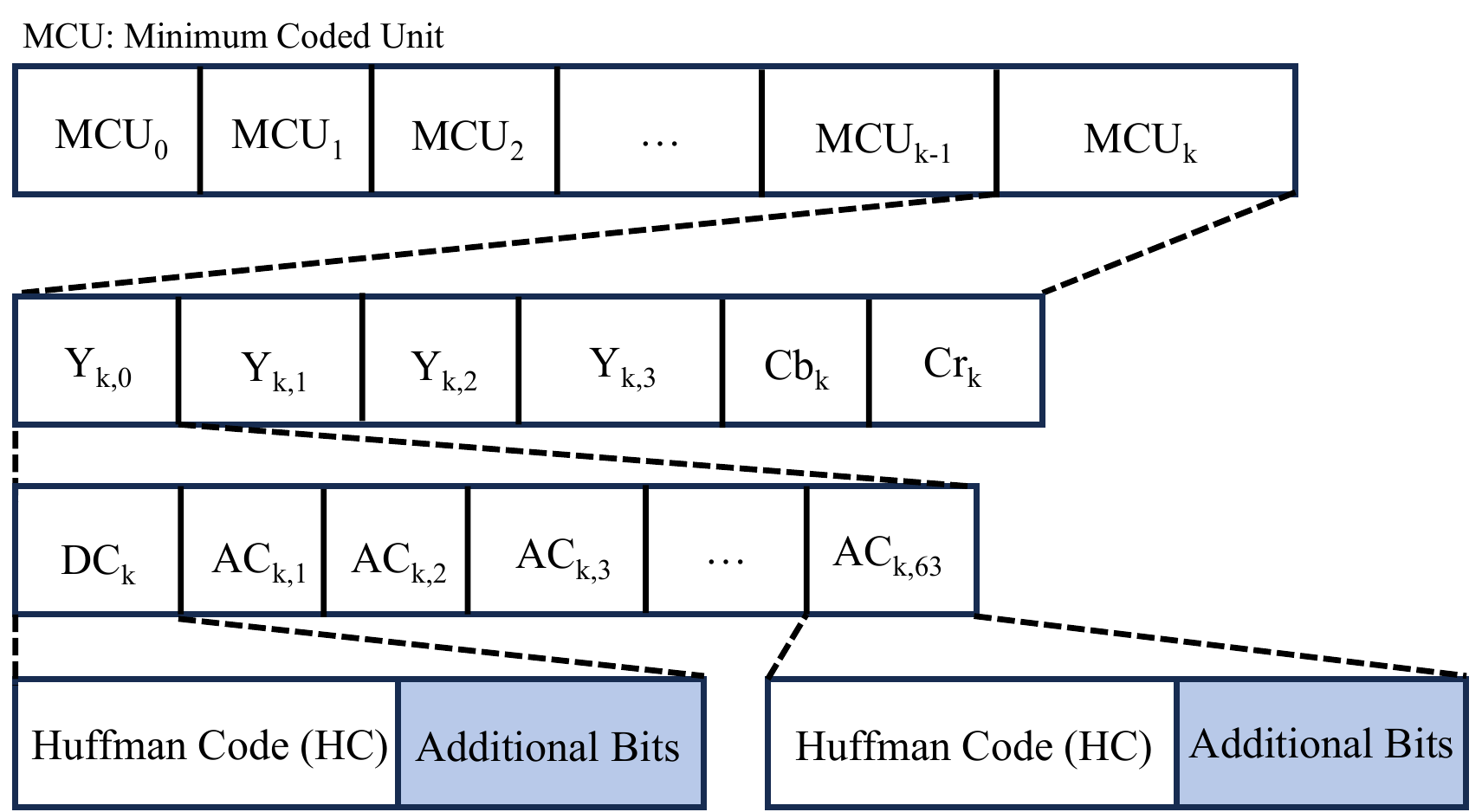}
\end{center}
\caption{\textcolor{black}{Structure} of image data.}
\label{fig:image_data}
\vspace*{-3pt}
\end{figure}

Fig. \ref{fig:image_data} \textcolor{black}{depicts} the \textcolor{black}{structure} of the image data of a JPEG bitstream for the case of 4:2:0 color subsampling. Image data is divided and processed in units called minimum coded units (MCUs). Each MCU \textcolor{black}{is composed} of four luminance components (Y) and two chroma components (Cb, Cr). Each component contains a DC coefficient and AC coefficients, and the DC coefficient is recorded as the difference value from the previous DC coefficient. Finally, these bit sequences are divided and stored byte by byte. Due to this process, FF$_{(16)}$ may occur in the image data. Therefore, the JPEG encoder inserts 00$_{(16)}$ immediately after FF$_{(16)}$ occurs in the data to \textcolor{black}{tell} it \textcolor{black}{apart} from the first byte of a marker code. Additionally, when the JPEG decoder detects FF00$_{(16)}$, it reads only FF$_{(16)}$ and skips 00$_{(16)}$. This process is called ``byte stuffing.''

One of the JPEG marker codes is called the RST marker, which can be \textcolor{black}{placed} at fixed intervals among MCUs. The restart interval ($RI$), which is the interval at which the RST marker is \textcolor{black}{placed}, is set during JPEG encoding. When a RST marker is not used, an error that \textcolor{black}{arises} in the DC coefficient in a bitstream propagates to the end of the bitstream. \textcolor{black}{In contrast}, the value of a DC coefficient \textcolor{black}{following} a RST marker is stored as its original value. Therefore, using RST markers prevents errors from propagating among DC coefficients in bitstreams.

\section{Security Evaluation}
\subsection{Bitstream-level JPEG Encryption with RST Markers} \label{3a}

\begin{figure*}[t]
\begin{center}
\includegraphics[width=180mm]{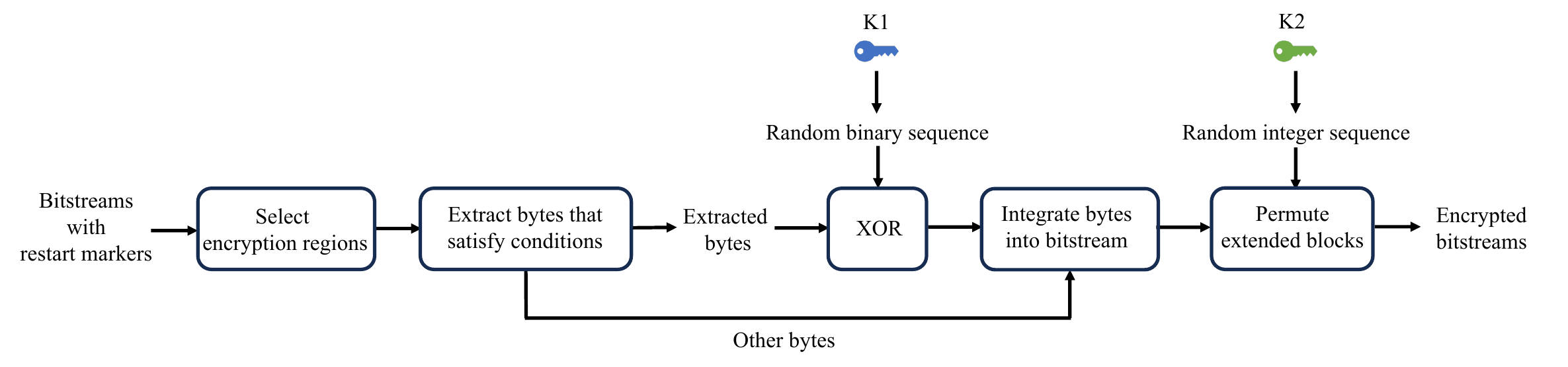}
\caption{Block diagram of encryption process.}
\label{fig:encryption_pro}
\end{center}
\end{figure*}

The bitstream-level JPEG encryption method with RST markers \cite{enc_rst} is summarized here. Fig. \ref{fig:encryption_pro} shows the encryption process of the method \cite{enc_rst}. The method utilizes RST markers for encryption, where a \textcolor{black}{sequence} of MCUs \textcolor{black}{divided} by RST markers is referred to as an extended block (see Fig. \ref{fig:rst}). The encryption procedure for this method is shown below.

\begin{figure}[t]
\begin{center}
\includegraphics[width=80mm]{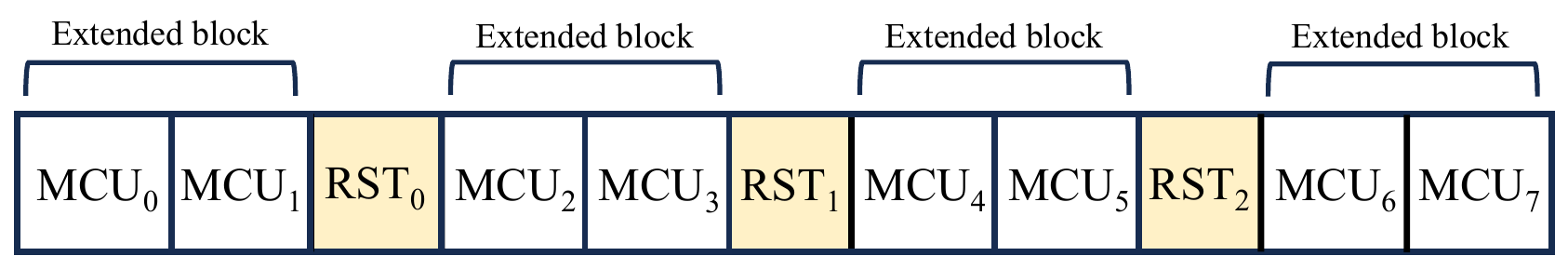}
\caption{Bitstreams after insertion of RST markers ($RI$ = 2).}
\label{fig:rst}
\end{center}
\end{figure}

\renewcommand{\labelenumi}{Step \theenumi:}
\begin{enumerate}
\setlength{\leftskip}{20pt}
\item	Select an $RI$ and prepare a bitstream including RST markers to be encrypted. 
\item	Select encryption regions.
\item	\textcolor{black}{Extract} bytes that satisfy encryption conditions from the extended blocks to maintain the same file size as that before encryption. 
\item   \textcolor{black}{Prepare} a binary sequence of pseudorandom numbers (PRNs) by using secret K1. \textcolor{black}{Perform} an exclusive-or (XOR) operation between additional bits of the \textcolor{black}{extracted} bytes and the sequence of PRNs.
\item   Replace the additional bits with the XOR results.
\item   Randomly permute the positions of the extended blocks defined in Step 1 with secret K2.
\end{enumerate}

The encryption conditions in Step 3 were discussed in \cite{pre, pre_jpn}, in which the bytes of a bitstream are categorized into five cases, as shown in Fig. 6, where the white, black, and red areas \textcolor{black}{denote} Huffman codes (HCs), additional bits, and bytes placed immediately after FF$_{(16)}$, respectively. Each \textcolor{black}{pattern} is defined below.

\begin{figure}[t]
\begin{center}
\includegraphics[width=80mm]{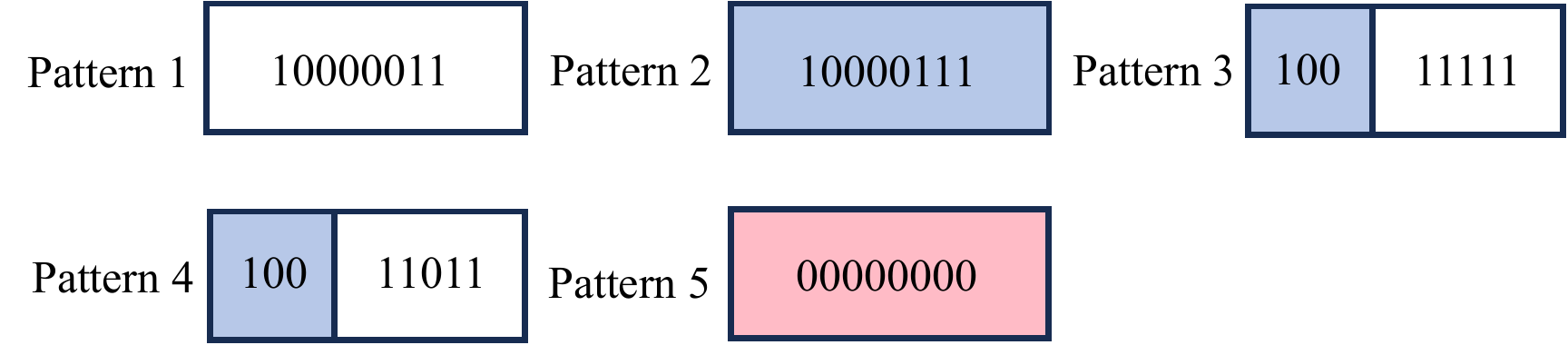}
\end{center}
\caption{Five types of bytes.}
\label{fig:cases}
\vspace*{-3pt}
\end{figure}

\renewcommand{\labelenumi}{\textcolor{black}{Pattern} \theenumi:}
\begin{enumerate}
\setlength{\leftskip}{20pt}
\item	Consists \textcolor{black}{solely} of HCs.
\item	Consists \textcolor{black}{solely} of additional bits.
\item	Consists of HCs and additional bits, and every bit in the HC is 1.
\item   Consists of HCs and additional bits, and the HC \textcolor{black}{contains} 0.
\item   Consists \textcolor{black}{solely} of 0, and the byte is placed immediately after FF$_{(16)}$.
\end{enumerate}

\noindent Among these \textcolor{black}{patterns}, only additional bits in \textcolor{black}{Pattern} 4 can be encrypted. 

\begin{figure*}[t]
\begin{center}
\begin{minipage}[b]{0.23\linewidth}
\begin{center}
\includegraphics[width=25mm]{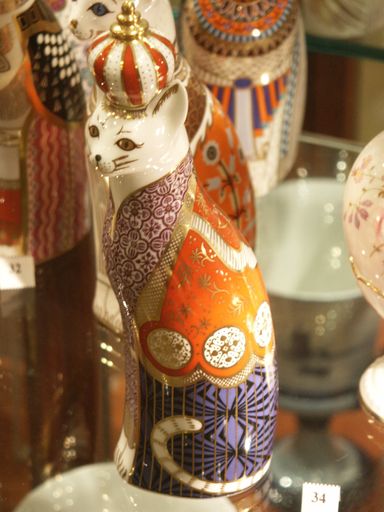}
\subcaption{Original}\label{fig:example_org}
\end{center}
\end{minipage} 
\begin{minipage}[b]{0.23\linewidth}
\begin{center}
\includegraphics[width=25mm]{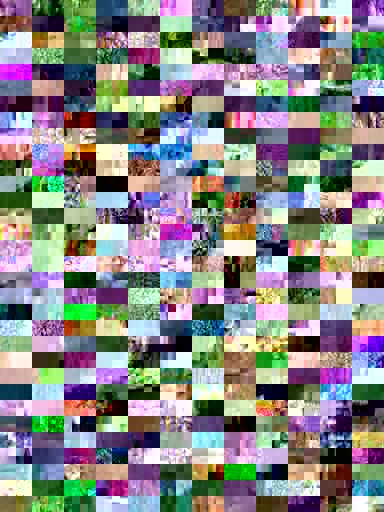}
\subcaption{Encrypted ($RI = $ 2)}
\end{center}
\end{minipage} 
\begin{minipage}[b]{0.23\linewidth}
\begin{center}
\includegraphics[width=25mm]{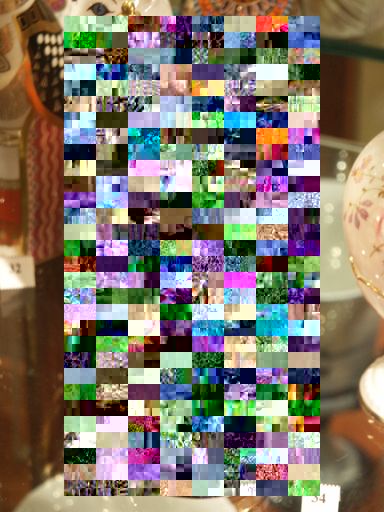}
\subcaption{Partially-encrypted ($RI = $ 2)}
\end{center}
\end{minipage} 
\begin{minipage}[b]{0.23\linewidth}
\begin{center}
\includegraphics[width=25mm]{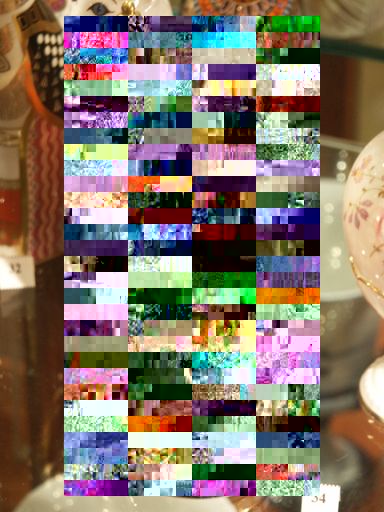}
\subcaption{Partially-encrypted ($RI = $ 4)}
\end{center}
\end{minipage}
\caption{Examples of encrypted images (ucid00459).}
\label{fig:example}
\end{center}
\end{figure*}

In \cite{enc_rst}, each extended block is encrypted separately. Thus, it is possible to generate a partially encrypted image that contains both unencrypted and encrypted regions. Fig. \ref{fig:example} illustrates an example of images encrypted with this method.

\subsection{Threat Model}
A threat model includes a set of assumptions, such as an adversary's goals, knowledge, and capabilities. The \textcolor{black}{aim} of an attacker is to restore visual information from encrypted data. We assume that the attacker is able to use encrypted data and the encryption algorithm but does not have the secret key. Accordingly, the attacker can only perform ciphertext-only attacks (COAs) using encrypted images.

\subsection{Security Analysis}
Several COAs have been studied to \textcolor{black}{restore} visual information from encrypted images \cite{enc1, enc4, enc6, pre_jpn}. In this paper, we use key space analysis, key sensitivity analysis, the non-zero counting attack (NZCA), and histogram analysis for security evaluation. The relationship between the use of RST markers and the attack resistance is discussed. 

\subsubsection{Key space analysis}
Key space is the total number of patterns that can be \textcolor{black}{produced} by a encryption algorithm. 
Here, we analyze the encryption method in the case where all DCT coefficients in the whole image are encrypted. The algorithm encrypts additional bits within bytes that satisfy specific conditions and permutes the positions of the extended blocks, as shown in \ref{3a}.
We \textcolor{black}{assume} that JPEG encoding is applied to an image with a size of $M \times N$ under $RI =$ r and 4:2:0 color subsampling.
First, considering that the number of bytes to be encrypted is $T$, the minimum and maximum sizes of the key space derived from the encryption of the additional bits $S_{enc\_min}$ and $S_{enc\_max}$ are \textcolor{black}{given by}

\begin{align}
  & S_{enc\_min} = 2^T, \\
  & S_{enc\_max} = 2^{7T}.
\end{align}

\noindent Note that the minimum number of bits to be encrypted is one, and the maximum is seven within a byte that satisfies the encryption conditions. The key space obtained from the permutation of the extended blocks, $S_{bp}$, is expressed as

\begin{align}
  & S_{bp} = \left\lfloor \left(\left\lceil \frac{M}{16} \right\rceil \times \left\lceil \frac{N}{16} \right\rceil \times \frac{1}{r}\right)\right\rfloor!. \label{eq3}
\end{align}

\noindent Accordingly, the overall minimum key space $S_{min}$ and maximum key space $S_{max}$ are given by
\begin{align}
   S_{min} &= S_{enc\_min} \times S_{bp} \notag \\ 
&= 2^T \times \left\lfloor \left(\left\lceil \frac{M}{16} \right\rceil \times \left\lceil \frac{N}{16} \right\rceil \times \frac{1}{r}\right)\right\rfloor!, \\
   S_{max} &= S_{enc\_max} \times S_{bp} \notag
  \\ &= 2^{7T} \times \left\lfloor \left(\left\lceil \frac{M}{16} \right\rceil \times \left\lceil \frac{N}{16} \right\rceil \times \frac{1}{r}\right)\right\rfloor!.
\end{align}

\noindent Equation (\ref{eq3}) shows that the number of extended blocks increases as $RI$ becomes shorter. Therefore, an encrypted image with a shorter $RI$ has a larger key space. The algorithm for the method with RST markers can use a shorter $RI$ and thus can provide a larger key space. For example, if the image in Fig \ref{fig:example_org} is encoded with $r$ = 4 as an $RI$, $T$ = 37,031 is given. Thus,
the minimum key space $S_{min}$ can be expressed by

\begin{align}
   S_{min} &= 2^{37,031} \times \left\lfloor \left(\left\lceil \frac{384}{16} \right\rceil \times \left\lceil \frac{512}{16} \right\rceil \times \frac{1}{4}\right)\right\rfloor! > 2^{256}.
\end{align}

\noindent This is sufficiently larger than $2^{256}$, which is the space of a key with 256 bits. If we use $r$ = 2, $S_{min}$ is
larger than in the case of $r$ = 4. Accordingly, the bitstream-level encryption method with RST markers has enough key space.

\subsubsection{Key sensitivity analysis}
Attack-resistant encryption methods should be sensitive even to a slight change in the key. We analyze key sensitivity by considering two cases:

\renewcommand{\labelenumi}{Case \theenumi:}
\begin{enumerate}
\setlength{\leftskip}{20pt}
\item	Encryption using an encryption key that differs from the original key by only one bit.
\item	Decryption using an incorrect key that differs from the correct key by only one bit.
\end{enumerate}

In Case 1, the encrypted image must be \textcolor{black}{wholly} different from the image encrypted with the original key. In Case 2, the image decrypted with an incorrect key must be \textcolor{black}{wholly} different from the original image.

\subsubsection{NZCA}
The sketch attack tries to obtain contour information of the original one from the encrypted image. We use the non-zero-counting attack (NZCA), which is a sketch attack for encrypted JPEG images \cite{enc1}. This attack attempts to restore the original image's contours on the basis of the number of non-zero coefficients in each MCU of the encrypted image. The use of RST markers will be evaluated to generate encrypted images robust against NZCA.

\subsubsection{Histogram analysis}
The histogram is very useful for an adversary to analyze a data distribution graphically in the encrypted domain. An adversary tries to analyze the distribution of pixel frequencies in encrypted images to apply attacks. Accordingly, attack-resistant encryption methods should provide encrypted images with histograms similar to those of other encrypted images.

\section{Experimental Results}
\subsection{Setup}
In experiments, we used 1,338 test images with 384 × 512 pixels or 512 × 384 pixels from Uncompressed Colour Image Database \cite{ucid}. JPEG compression was performed using libjpeg \cite{jpeg}, with the color subsampling set to 4:2:0 and the quality factor Q set to 80. Binary sequences of PRNs were \textcolor{black}{prepared} using HMAC\_DRBG \cite{nist} with a 384-bit key. For all images, we encrypted the \textcolor{black}{whole} image and permuted the positions of all extended blocks.

\subsection{Protection of Visual Information}
\begin{figure*}[t]
\begin{center}
\begin{minipage}[b]{0.45\linewidth}
\begin{center}
\includegraphics[width=70mm]{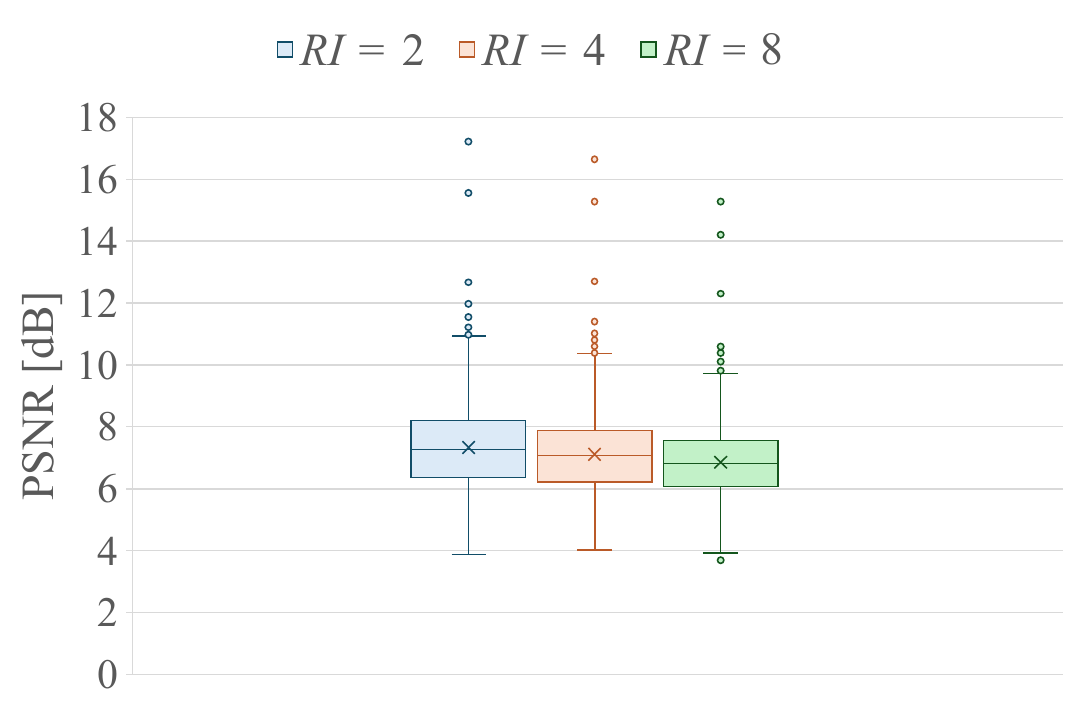}
\subcaption{PSNR}\label{fig:PSNR_enc}
\end{center}
\end{minipage} 
\begin{minipage}[b]{0.45\linewidth}
\begin{center}
\includegraphics[width=70mm]{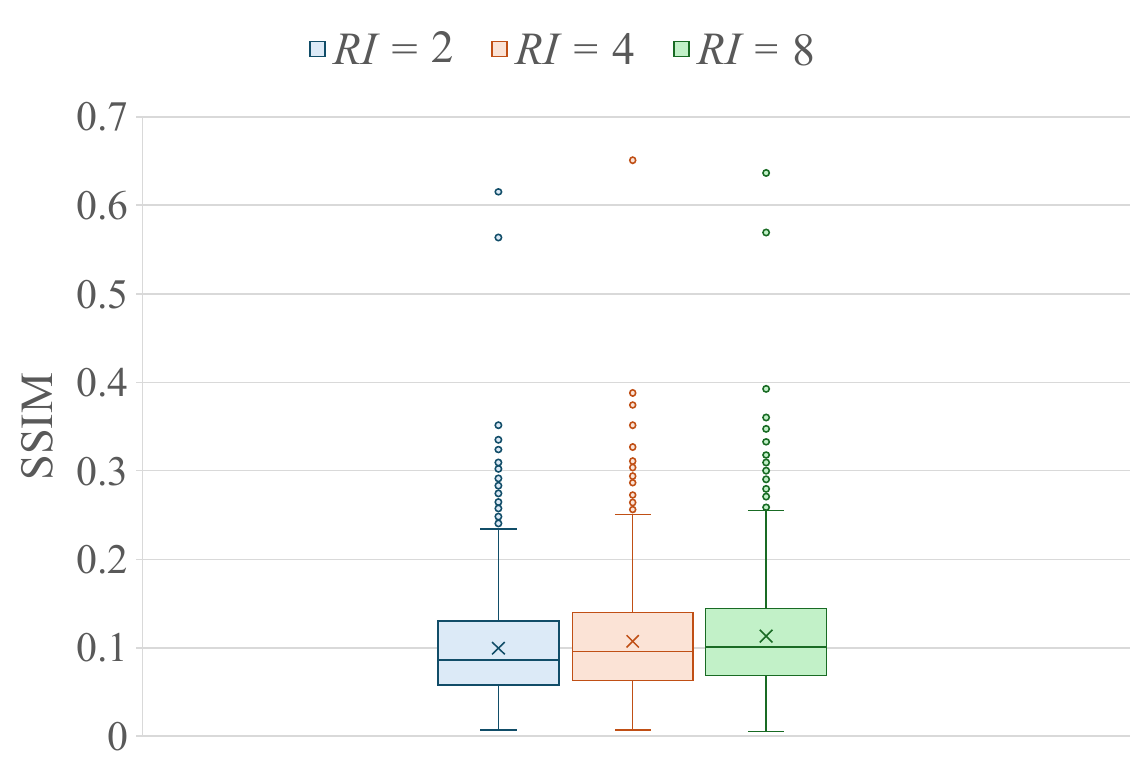}
\subcaption{SSIM}\label{fig:SSIM_enc}
\end{center}
\end{minipage}
\caption{Evaluation of visual protection. Top of box, bottom of box, and line in middle represent 75th, 25th, and 50th percentile, respectively. Whiskers denote highest and lowest values that are not outliers. Circles are outliers, and crosses are means.}
\label{fig:enc_box}
\end{center}
\end{figure*}

The encryption method with RST markers can permute the positions of blocks \textcolor{black}{at random}, thereby strongly protecting the visual information of images, compared with previous methods without block permutation \cite{pre, pre_jpn, enc2}. For objectively evaluating the protection strength, peak signal-to-noise ratio (PSNR) and structural similarity (SSIM) values of encrypted images are shown as boxplots in Fig. \ref{fig:enc_box}. From the figure, it is confirmed that the method generates encrypted images without identifiable visual information.

\subsection{Key Sensitivity}

\begin{figure*}[t]
\begin{center}
\begin{minipage}[b]{0.45\linewidth}
\begin{center}
\includegraphics[width=70mm]{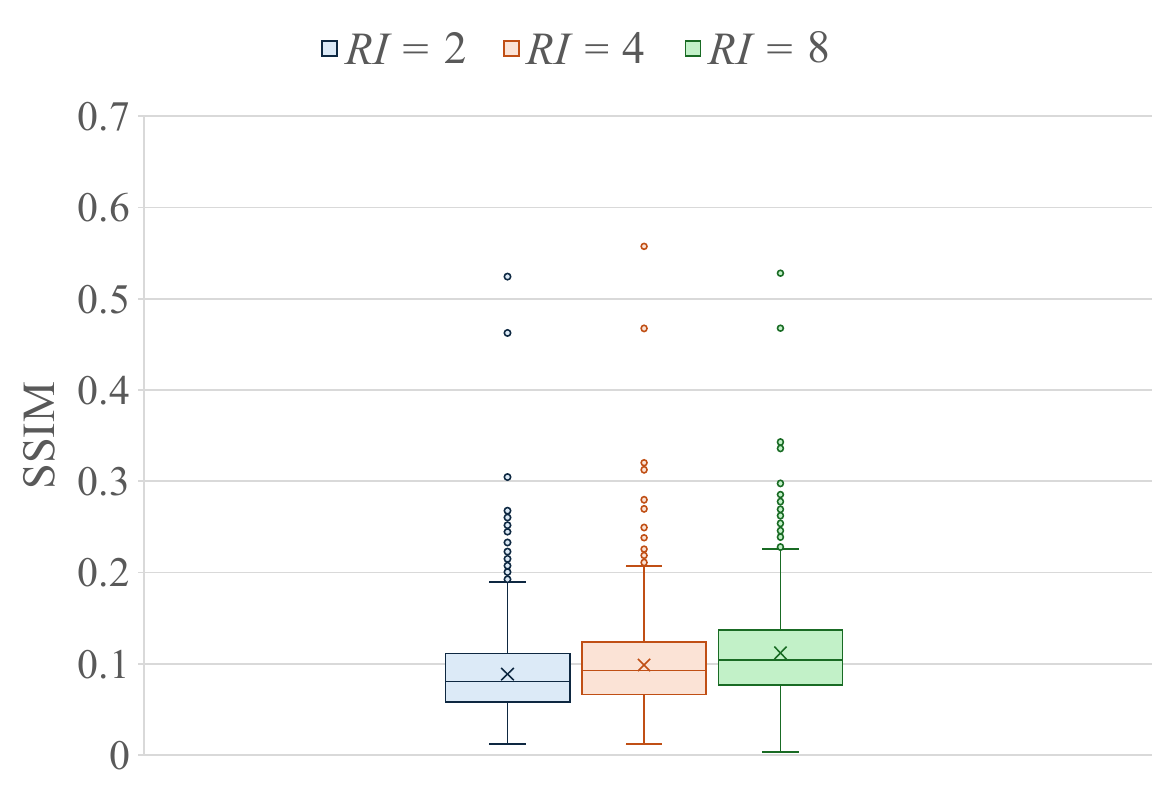}
\subcaption{Encryption with different keys (Case 1)}\label{fig:SSIM_enc_dif}
\end{center}
\end{minipage} 
\begin{minipage}[b]{0.45\linewidth}
\begin{center}
\includegraphics[width=70mm]{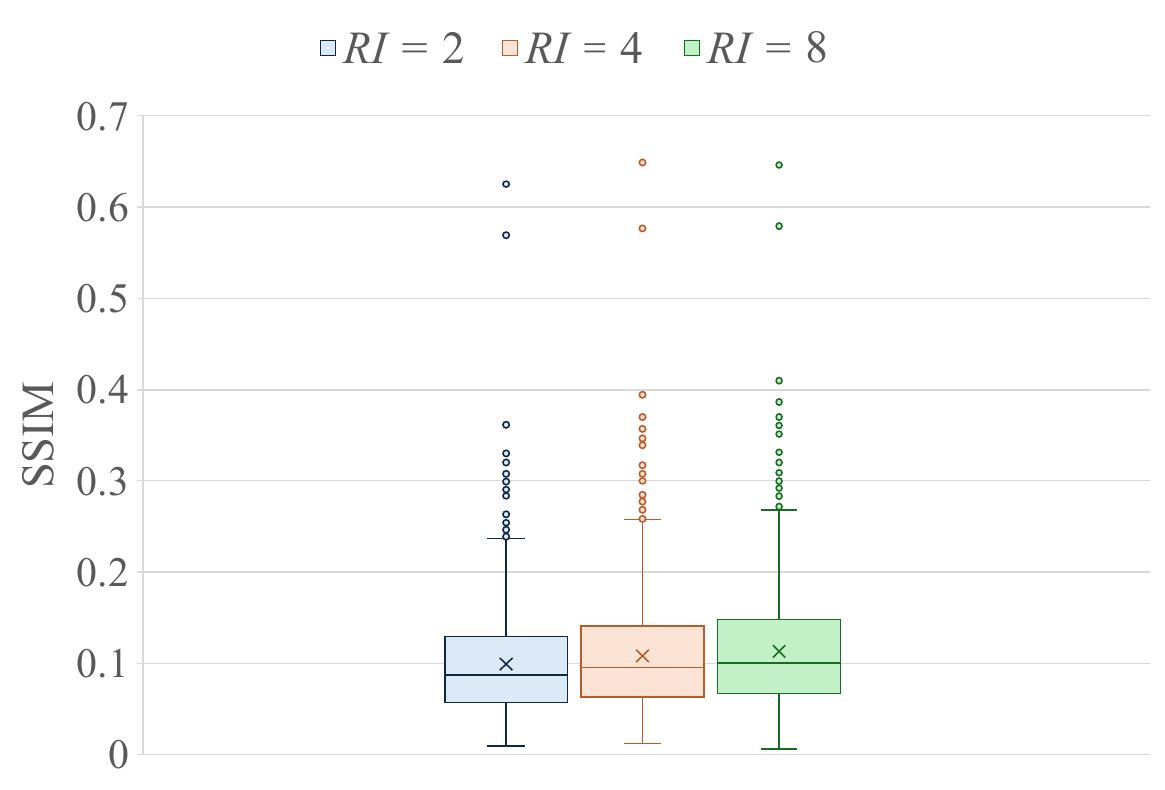}
\subcaption{Decryption with incorrect keys (Case 2)}\label{fig:SSIM_dec_dif}
\end{center}
\end{minipage}
\caption{Key sensitivity analysis results. Top of box, bottom of box, and line in middle represent 75th, 25th, and 50th percentile, respectively. Whiskers denote highest and lowest values that are not outliers. Circles are outliers, and crosses are means.}
\label{fig:key_dif}
\end{center}
\end{figure*}

Fig. \ref{fig:key_dif} illustrates the results of the key sensitivity analysis, where the SSIM values indicate the difference between two images. In Fig. \ref{fig:SSIM_enc_dif}, two images were encrypted with keys that differed by a single bit. In contrast, in Fig. \ref{fig:SSIM_dec_dif}, the two images were decoded from an encrypted one by using the correct key and an incorrect key that differed from the correct key by a single bit. From the figure, the method was verified to have sufficiently high key sensitivity.  

\subsection{NZCA}
\begin{figure*}[t]
\begin{center}
\includegraphics[width=180mm]{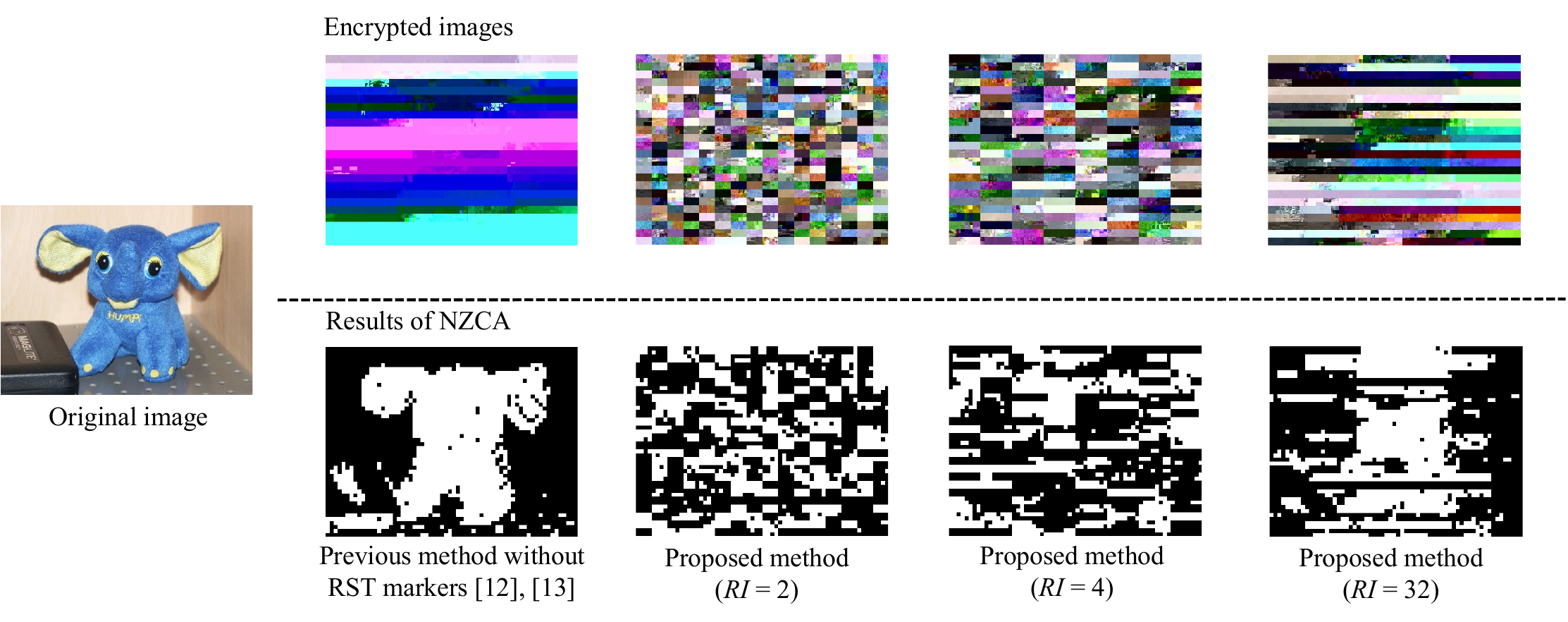}
\end{center}
\caption{Results of non-zero-counting attack (ucid00294).}
\label{fig:nzca}
\vspace*{-3pt}
\end{figure*}

NZCA was \textcolor{black}{carried out} on the luminance component of an encrypted image. As illustrated in Fig. \ref{fig:nzca}, in the previous method without RST markers \cite{pre, pre_jpn}, the original image was \textcolor{black}{revealed} in its contour. In \textcolor{black}{contrast}, the method with RST markers concealed the original image. This was \textcolor{black}{attained} by replacing extended blocks. However, when a large $RI$ was adopted, the outline of the original image was revealed in some areas. Accordingly, the use of RST markers is important for enhancing robustness against attacks, but the value of $RI$ should be carefully selected.

\subsection{Histogram Analysis}
\begin{figure*}[t]
\begin{center}
\includegraphics[width=190mm]{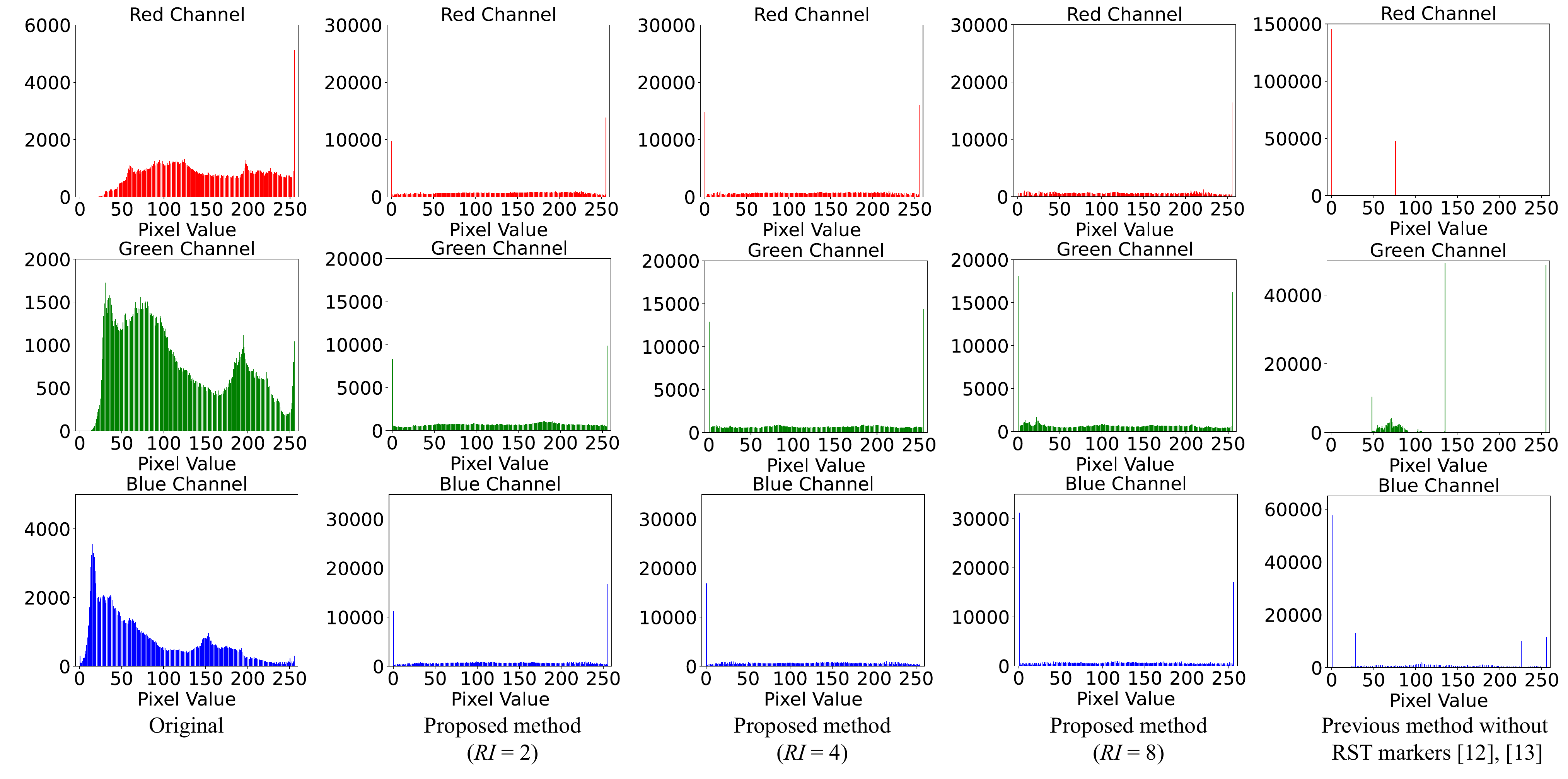}
\end{center}
\caption{Histogram analysis (ucid00459).}
\label{fig:hist}
\end{figure*}

Fig. \ref{fig:hist} illustrates the R, G, and B histograms of the original ucid00459 image, which is shown in Fig. \ref{fig:example_org}, and its encrypted images. In the figure, there are histograms for four encrypted images.
Three of them were derived by using our method with RST markers, and the other was derived by using the previous method. In our method with RST markers, $RI$ was defined as 2, 4, and 8. The images encrypted using this method were compared with the original image and those encrypted with the previous method. Attack-resistant encryption methods should provide an encrypted image with a totally different histogram from that of the original image and analogous histograms among the three color channels. From the figure, we can see that the R, G, and B channels in the encrypted image have analogous histograms. In addition, the use of a small $RI$ can strongly reduce the identifying characteristics of the image. Other images were verified to have similar trends.

\section{Conclusion}
In this paper, we conducted security analyses on a state-of-the art encryption method with RST markers against ciphertext-only attacks. It was confirmed that the RST markers used in the method contribute to enhancing security in terms of key space, key sensitivity, NZCA, and histogram analysis. As a result, the method was demonstrated not only to outperform conventional JPEG encryption methods in some important respects but also to maintain a high security level. Furthermore, it was shown that the attack resistance varies with the length of the restart interval.

\section*{Acknowledgment}
This work was partially supported by JSPS KAKENHI Grant Number JP21H01327.


\end{document}